\documentclass[journal]{IEEEtran}

\usepackage{graphicx}
\usepackage{amsmath}
\usepackage{array}
\usepackage{cite}
\usepackage[caption=false,font=footnotesize]{subfig}
\usepackage{dblfloatfix}
\usepackage{url}
\usepackage{booktabs}
\usepackage{float}
\usepackage{xcolor}
\usepackage{pifont}
\usepackage{amssymb}

\newcommand{\Rs}{$R$\textsubscript{s}}
\newcommand{\Rsdc}{$R$\textsubscript{s, DC}}
\newcommand{\Tc}{$T$\textsubscript{c}}
\newcommand{\ns}{$n$\textsubscript{s}}
\newcommand{\Lk}{$L$\textsubscript{k}}

\begin{document}

\title{Contactless terahertz mapping of wafer-scale superconducting NbTiN thin films}

\author{Yayi~Lin, Marc~Neis, Marcello~Pio~Guardascione, Janine~Lorenz, Thomas~J.~Smart, F.~Stefan~Tautz, Felix~Lüpke, Frederik~Bolle, Martin~Dressel, Rami~Barends, Pavel~A.~Bushev, and Marc~Scheffler
\thanks{This work was supported by the German Federal Ministry of Research, Technology and Space (BMFTR) within the project QSolid (FKZ: 13N16149 and 13N16159) and by the Carl-Zeiss-Stiftung (QPhoton). J.L., F.S.T. and F.L. acknowledge funding from the Bavarian Ministry of Economic Affairs, Regional Development and Energy within Bavaria’s High-Tech Agenda Project ”Bausteine für das Quantencomputing auf Basis topologischer Materialien mit experimentellen und theoretischen Ansätzen”. F.L. further acknowledges funding by the Deutsche Forschungsgemeinschaft (DFG, German Research Foundation) within the Emmy Noether Programme (project no. 511561801). \textit{(Corresponding author: Marc~Scheffler.)}}%
\thanks{Yayi~Lin, Frederik~Bolle, Martin~Dressel, and Marc~Scheffler are with the 1.Physikalisches Institut, Universität Stuttgart, 70569 Stuttgart, Germany (e-mail: yayi.lin@pi1.uni-stuttgart.de, marc.scheffler@pi1.uni-stuttgart.de).}%
\thanks{Martin~Dressel and Marc~Scheffler are also with the Center for Integrated Quantum Science and Technology (IQST), Universität Stuttgart, 70569 Stuttgart, Germany}%

\thanks{Marc~Neis, Marcello~Pio~Guardascione, Rami~Barends, and Pavel~A.~\mbox{Bushev} are with the Institute for Functional Quantum Systems (PGI-13), Forschungszentrum Jülich, 52425 Jülich, Germany}%
\thanks{Marc~Neis, Marcello~Pio~Guardascione, and Rami~Barends are also with the Department of Physics, RWTH Aachen University, 52074 Aachen, Germany}%

\thanks{Janine~Lorenz, F.~Stefan~Tautz, and Felix~Lüpke are with the Peter Grünberg Institute (PGI-3), Forschungszentrum Jülich, 52425 Jülich, Germany}%
\thanks{Janine~Lorenz and F.~Stefan~Tautz are also with the Institute of Experimental Physics IV A, RWTH Aachen Universität, 52074 Aachen, Germany}%

\thanks{Thomas~J.~Smart is with the Peter Grünberg Institute (PGI-9), Forschungszentrum Jülich, 52425 Jülich, Germany}%

\thanks{Janine Lorenz, F. Stefan Tautz, Felix Lüpke, and Thomas J. Smart are also with Jülich Aachen Research Alliance (JARA), Fundamentals of Future Information Technology, 52425 Jülich, Germany}%

\thanks{Felix~Lüpke is also with the Institute of Physics II, Universität zu Köln, 50937 Köln, Germany}%
}

\maketitle


\begin{abstract}
For large-scale superconducting quantum technology, e.g.\ quantum computing,  the homogeneity of wafer-scale superconducting thin films is vital for consistent performance of the fabricated devices. Tera\-hertz (THz) spectroscopy as a contactless and non-destructive measurement technique is a powerful tool to characterize the superconducting films. In this work, a set of niobium titanium nitride (NbTiN) thin films on 4-inch and 6-inch silicon wafers, grown via plasma-enhanced magnetron sputtering, are investigated via THz spectroscopy: full wafers are mapped at room temperatures and exemplary segments are characterized at cryogenic temperatures. The deviations in observed sheet resistance depend on the used deposition device and the film thickness. While the deviations in superconducting sheet kinetic inductance match those of the normal-state sheet resistance, the critical temperature and energy gap exhibit little variation. This THz mapping technique demonstrates the feasibility of evaluating wafer-scale superconducting thin films before lithography, facilitating preparation of the thin films for reproducible device fabrication.
\end{abstract}



\section{Introduction}

\IEEEPARstart{F}{ollowing} the emergence of quantum mechanics, quantum computing, as a new concept, has gradually revealed the advantages arising from quantum entanglement beyond the capability of classical bits~\cite{1982feynman, 2025oukaira, 2019arute}. Currently, various quantum computing architectures are extensively studied~\cite{2010ladd, 2021leon}, including but not limited to, atomic systems~\cite{2022graham, 2025foss}, photonic systems~\cite{2026zhu}, and solid-state systems~\cite{2026tosato, 2020kjaergaard}. Specifically, as one of the most promising solid-state platforms, superconducting thin films are widely employed to realize quantum processors~\cite{2004devoret, 2008clarke, 2017wendin, 2019krantz, 2020kjaergaard, 2021siddiqi, 2025abughanem, 2025jiang} as well as low-noise amplifiers~\cite{2015macklin, 2021esposito, 2023klimovich}, partly due to its compatibility with well-established CMOS-type fabrication technologies, enabling on-wafer integration~\cite{2008schoelkopf, 2020kjaergaard, 2019krantz, 2024damme, 2025potjan}. Other established and continuously evolving fields of application of superconducting thin-film devices include various kinds of detectors such as kinetic inductance detectors (KIDs)~\cite{2003day, 2022baselmans, 2014Bueno} and superconducting nanowire single-photon detectors (SNSPDs)~\cite{2021esmaeil, 2023chen}.

At the material level, several key superconducting parameters affect the performance of these quantum devices, such as critical temperature \Tc, superconducting energy gap 2$\Delta_\mathrm{0}$, superfluid density \ns\textsubscript{0}, sheet kinetic inductance \Lk, and quasiparticle excitations. 
For large-scale device fabrication~\cite{2024kim}, these parameters should be as homogeneous as possible at the wafer scale, and there should be techniques available to quantify this homogeneity~\cite{2017brun, 2022west, 2024zhao, 2026lorenz}.
One of the first steps in characterizing superconducting thin films usually addresses DC properties like resistivity $\rho$, residual resistivity ratio (RRR), and \Tc~via four-point transport measurement with electrical contacts~\cite{2003Matsunaga}.
In the field of quantum computing, RF properties are of upmost importance such that the following characterization steps often involve the fabrication and characterization of microwave resonators, which gives insight into the RF response, quasi-particle losses, and \Lk. All of them are important characteristics for expected quantum device performance~\cite{2016goetz, 2024kim, 2026smart}.
These techniques can also be used to investigate wafer-scale homogeneity of superconducting thin films~\cite{2017Thoen}, but they typically require a measurement of multiple devices per wafer after some kind of contacting or lithographic processing.
In the present work, we investigate a different approach, namely wafer-scale THz mapping. This technique probes the electrodynamic response of the materials in the relevant high-frequency regime, and thus delivers key information for device performance without the need of lithography.

As a quasi-optical beam, the THz radiation can be used to study the materials' properties in a contactless manner by detecting the transmission or reflection. Because the frequency range covers 0.1--3\,THz, where the superconducting features typically appear, THz spectroscopy is particularly suitable for studying superconducting materials. Recently, various types of spectroscopy based on different THz sources and detectors are widely used. As representative examples, THz frequency-domain spectroscopy (THz-FDS) uses a continuous-wave source, such as a set of backward wave oscillators allowing a frequency range between 30\,GHz to 1.5\,THz, and thermal detectors such as a Golay cell or bolometer. Frequency-dependent transmission is observed directly and used to determine the sheet resistance and superconducting parameters~\cite{1996pronin, 2013pracht, 2016pracht, 2025saritas}. Meanwhile, THz time-domain spectroscopy (THz-TDS) with a slightly higher frequency range (300\,GHz--3\,THz) is based on the Fourier transform of an ultra-short THz pulse~\cite{2011jepsen, 2012lloyd}. It is commonly used to investigate superconductors with \Tc~in the range of 10\,K or higher~\cite{2023lee, 2003brucherseifer, 2010valdesaguilar, 2023Khan}.

\begin{table*}
    \centering
    \begin{tabular}{|c|c|c|c|c|c|c|c|c|c|c|}
    \hline
        Wafer No. & wC & wD & wF & wH & wI & wJ & wK & wL & wM & wN \\
        \hline
        Wafer diameter & 4" & 4" & 6" & 4" & 4" & 4" & 4" & 4" & 4" & 4" \\
        \hline
        Deposition device & A & A & B & B & B & B & B & B & B & B \\
        \hline
        Thickness $d$ & 50\,nm & 50\,nm & 50\,nm & 60\,nm & 50\,nm & 40\,nm & 30\,nm & 20\,nm & 10\,nm & 5\,nm \\
        \hline
        Average \Rs~($\Omega$/sq) & 27.36 & 26.82 & 20.31 & 18.56 & 22.85 & 29.44 & 39.25 & 62.79 & 146.03 & 375.10  \\
        \hline
        Std. of $\delta$\Rs & 1.77\% & 2.50\% & 2.66\% & 0.65\% & 1.23\% & 0.54\% & 0.59\% & 2.68\% & 4.95\% & 7.78\% \\
        \hline
    \end{tabular}
    \caption{Overview of investigated NbTiN films on Si wafers: wafer diameter in inches; device used for the NbTiN deposition; thickness $d$ of NbTiN thin film; average sheet resistance \Rs~from room-temperature wafer-scale mapping; standard deviation $\delta$\Rs~from Gaussian fit of the \Rs~distribution in Fig.~\ref{fig3_dev.mapping}.}
    \label{tab_info}
\end{table*}

In this work, two types of THz spectroscopy are employed to the wafer-scale investigation on niobium titanium nitride (NbTiN) superconducting thin films. Specifically, THz-FDS is used to demonstrate room-temperature wafer maps of the sheet resistance \Rs. Afterwards, low-temperature measurements via THz-TDS for the superconducting state of three pieces diced from different positions of the wafer are conducted. This study shows that in the normal state, the \Rs~homogeneity of these NbTiN wafers depends on the thin-film thickness $d$ and the used deposition device. In the superconducting state, while the \Tc~and the 2$\Delta_\mathrm{0}$ are quite uniform across the representative wafer, the \Lk~shows a deviation consistent with that of \Rs. Such a fast, contactless, and non-destructive technique in wafer scale opens novel strategies for thin-film quality screening for large-scale superconducting devices.

\begin{figure}
    \centering
    \includegraphics[clip, trim=0cm 0cm 0cm 0cm, width=0.99\columnwidth]{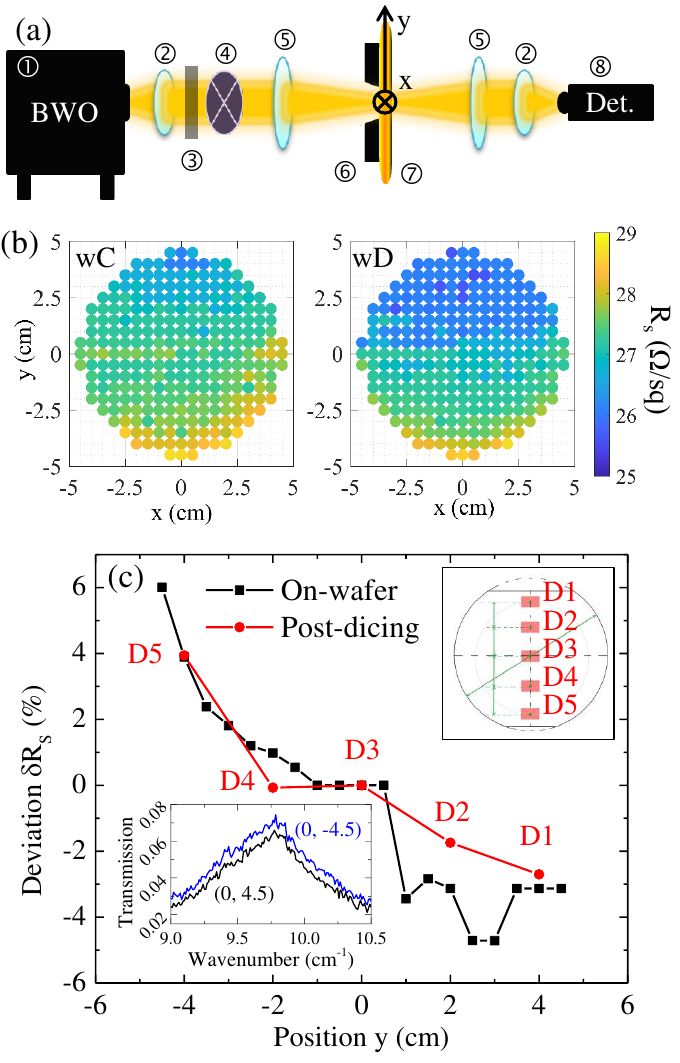}
    \caption{Overview of room-temperature mapping in THz-FDS. (a) THz-FDS setup. \ding{172} Backward wave oscillator (BWO), a continuous-wave THz source in the frequency range 8--13\,cm\textsuperscript{-1}, \ding{173} Teflon THz lenses to collimate output THz beam, \ding{174} Attenuator to avoid detector saturation, \ding{175} chopper to modulate THz beam, \ding{176} lenses to focus THz beam on the sample, \ding{177} sample holder with 6-mm diameter aperture, \ding{178} the measured wafer with NbTiN thin film, and \ding{179} Golay cell as THz detector. (b) \Rs~maps of two 4-inch wafers, wC (left) and wD (right). (c) \Rs~deviation for positions along y-axis with respect to the center D3 of wafer wD. The black data are from the on-wafer measurements shown in (b), while the red data are measured after dicing the wafer into smaller samples. The lower-left inset shows the raw THz transmission spectra of on-wafer data points at y = 4.5\,cm (black) and y = -4.5\,cm (blue). The upper-right inset indicates the positions of the diced pieces on the wafer.}
    \label{fig1_FDS_setup}
\end{figure}

\section{Sample preparation and measurement methods}

As one of the popular materials for superconducting quantum devices~\cite{2021esposito, 2025jiang, 2012Driessen}, superconducting thin-film NbTiN, whose 2$\Delta_\mathrm{0}$ locates in the THz regime~\cite{2013hong, 2023Khan}, is selected here to demonstrate the proposed mapping method. These films are grown on 525\,$\mu$m undoped silicon (Si) substrates (Siegert Wafers, $\rho>$10\,k$\Omega$cm) by plasma-enhanced reactive magnetron sputtering~\cite{2016bos}. Ten wafers are investigated to compare the influence of the wafer diameter, the deposition devices, and the thickness of the film. The specifications of these wafers are listed in Table~\ref{tab_info}. The wafers wC and wD were deposited in a Plassys Bestek sputter device for sample sizes up to 4\,inches (device A), while the wF to wN wafers were grown in a larger device designed for deposition on wafers with a diameter up to 6\,inches from the same company (device B). Since the two deposition devices differ in multiple ways, such as chamber geometry and distance from the target to the substrate, the growth parameters are adjusted to compensate for this in order to create comparable films. The growth parameters used for devices A and B are deposition power (750\,W and 900\,W), nitrogen flow (8\,sccm and 10\,sccm), deposition time, and rotation speed (5\,rpm and 10\,rpm), respectively.

\begin{figure}
    \centering
    \includegraphics[clip, trim=0cm 0cm 0cm 0cm, width=0.98\columnwidth]{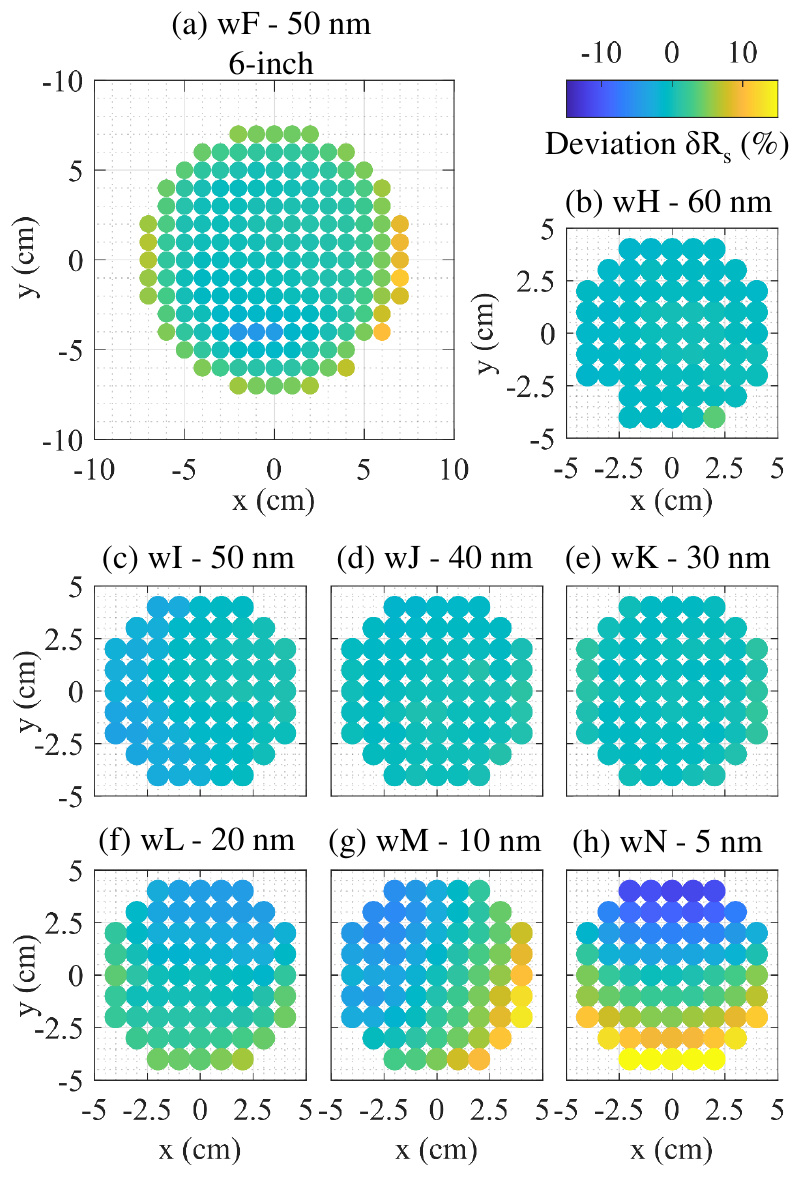}
    \caption{Percentage \Rs~deviation wafer-scale maps of (a) one 6-inch wafer wF and (b--h) seven 4-inch wafers wH--wN with different NbTiN film thicknesses between 60\,nm to 5\,nm. The deviations of \Rs~are indicated with respect to the center of each wafer.}
    \label{fig2_8_wafermappings}
\end{figure}

The room-temperature mapping is conducted in THz-FDS as shown in Fig.~\ref{fig1_FDS_setup}(a) by measuring the transmission of the sample for a grid of 6-mm-diameter spots. The transmission of a THz wave in the frequency range 9--10.5\,cm\textsuperscript{-1} is measured. Illustrative raw transmission spectra of on-wafer measurement are shown in the lower-left inset of Fig.~\ref{fig1_FDS_setup}(c). By using the Drude model, the sheet resistance \Rs~ is extracted while assuming that the scattering rate is much larger than that in the THz regime~\cite{2013pracht}. The \Rs~maps of the wafers wC and wD in Fig.~\ref{fig1_FDS_setup}(b) use a the spatial step between the points of 5\,mm, while a step of 10\,mm is used for the other wafers in Fig.~\ref{fig2_8_wafermappings}.

To characterize the superconducting properties, wafer wD is diced into several 10 $\times$ 11\,mm\textsuperscript{2} pieces in order to fit within the sample holder of the cryostat for low-temperature characterization \cite{2013pracht}. The layout of the diced pieces is shown in the upper-right inset of Fig.~\ref{fig1_FDS_setup}(c). Three diced pieces D1, D3, and D5 across the wafer are selected. First, the van der Pauw method is used to obtain the \Tc~performed in a homemade glass cryostat setup~\cite{2025Deshpande}. Next, a commercial THz spectrometer, TeraPulse 4000 (TeraView), is employed to measure the superconducting energy gap and superfluid density~\cite{2026bolle}. As shown in Fig.~\ref{fig4_TDSsetup_sigma}(a), the picosecond THz pulse is generated from a photo-conductive antenna based on a gallium-arsenide (GaAs) substrate, with an excitation laser pulse and applied bias voltage. It is focused on the sample by custom-made Teflon THz lenses. The receiver is also a GaAs-based antenna which outputs an electronic signal when the laser pulse is incident after the delay stage. The THz optical path is located within a nitrogen-purged chamber to avoid absorption of THz radiation by atmospheric water vapor. The samples are cooled in a homemade \textsuperscript{4}He cryostat with THz windows made by Mylar foils~\cite{2013pracht}. The base temperature of 1.5\,K is achieved via pumping on the liquid helium. By Fourier transforming the THz pulse, the frequency-dependent THz signals $\tilde{E}_\mathrm{s}$ of the sample and $\tilde{E}_\mathrm{r}$ of the bare reference substrate are used to calculate the optical conductivity $\tilde{\sigma}$ of the thin film by the Tinkham formula~\cite{2018neu}
\begin{equation}
    \frac{\tilde{E}_{s}(\nu)}{\tilde{E}_{r}(\nu)} = \frac{[1+\tilde{n}_{sub}(\nu)] \exp[i\psi(\nu)]}{1+\tilde{n}_{sub}(\nu)+Z_{0}\tilde{\sigma}(\nu)d},
    \label{TinkhamFormula}
\end{equation}
where $\nu$ is the wavenumber, $\tilde{n}_{sub}$ = 3.44 is the complex refractive index of the Si substrate, $Z_{0}$ is the vacuum impedance, $d$ is the thickness of the NbTiN thin film, and $\psi(\nu) = 2\pi\nu\tilde{n}_{sub}(\nu)\Delta d/c$ is the phase delay arising from the thickness difference $\Delta d$ between the bare reference substrate and the substrate with film, while $c$ is the speed of light in vacuum. In this work, $\Delta d$ = 0 is assumed for all samples.

\section{Result and discussion}

\subsection{Room-temperature \Rs~mapping}
For all of the ten NbTiN wafers, room-temperature mapping is demonstrated using THz-FDS. Here, one advantage compared to THz-TDS is the easy operation of variable wafer position without requiring a nitrogen purged chamber. As shown in the lower-left inset of Fig.~\ref{fig1_FDS_setup}(c), a Fabry--P\'erot oscillation arising from the dielectric substrate is observed within the measured frequency range. By translating the wafer to access different positions, the \Rs~of this certain area is obtained using the Drude model by fitting the Fabry--P\'erot oscillation. The \Rs~maps of the two wafers wC and wD are shown in Fig.~\ref{fig1_FDS_setup}(b). \Rs~at the center of wafers wC and wD is 27.21\,$\Omega$ and 26.95\,$\Omega$ and it deviates across the wafers by 2.59\,$\Omega$ and 3.11\,$\Omega$ (9.5\% and 11.5\%) respectively. The percentage deviation $\delta$\Rs\textsubscript{, (x,y)} with respect to \Rs\textsubscript{, (0,0)} at the wafer center (0,0) is defined as
\begin{equation}
    \delta R_{s, (x, y)} = \frac{R_{s, (x, y)}-R_{s, (0, 0)}}{R_{s, (0, 0)}}\times100~(\%), 
    \label{eq.deviation}
\end{equation}
where \Rs\textsubscript{, (x,y)} is the sheet resistance at position (x, y). Fig.~\ref{fig1_FDS_setup}(c) plots $\delta$\Rs\textsubscript{, (0,y)} along the vertical direction (x = 0) of wafer wD. The consistent deviation between the black (on-wafer) and red (post-dicing) data indicates the reliability of the measurement. 

\begin{figure}
    \centering
    \includegraphics[clip, trim=0cm 0cm 0cm 0cm, width=0.99\columnwidth]{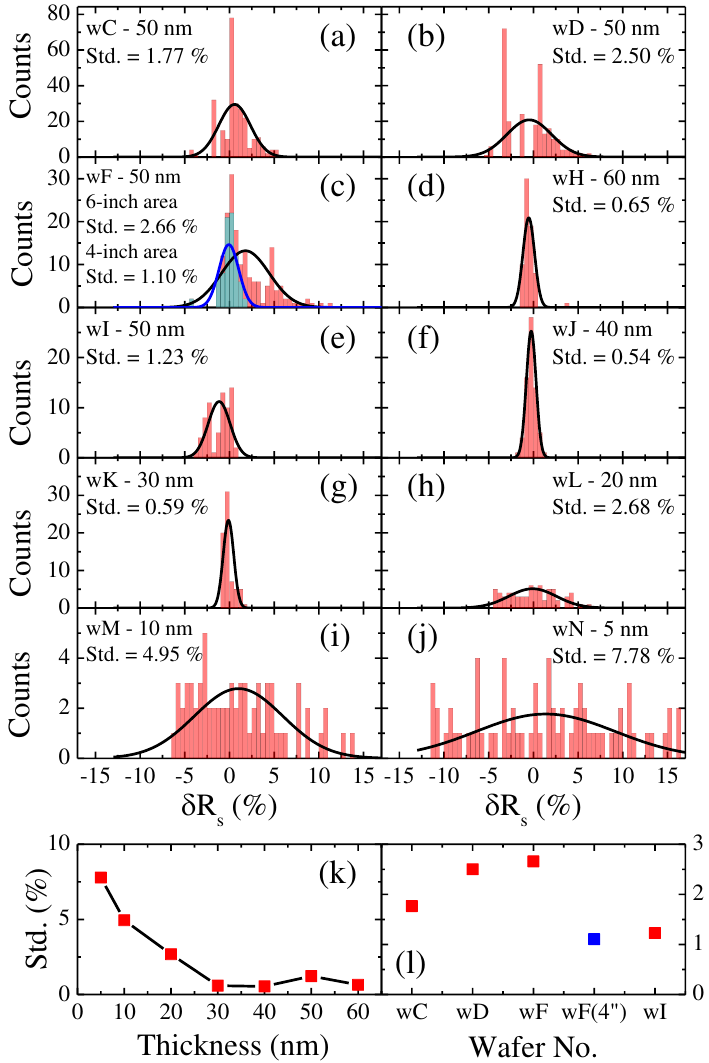}
    \caption{(a--j) The $\delta$\Rs~histograms of the wafers shown in Table~\ref{tab_info}. The solid lines are Gaussian fits and the fit width represents the standard deviation (Std.) of $\delta$\Rs. In (c), the red histogram and black line indicate the $\delta$\Rs~across the full 6-inch wafer wF, while the cyan histogram and blue line concern the smaller, center 4-inch area of the same wafer wF. (k, l) is the Std. obtained from the Gaussian fit. (k) is the standard deviation (Std.) as a function of film thickness, relating to (d--j). (l) is Std. for different wafers with 50-nm thin films, where the blue dot is related to the cyan and blue data in (c) and the red dots are for Std. across the full wafers.}
    \label{fig3_dev.mapping}
\end{figure}

The other eight wafers grown in the second deposition device (device B) are analyzed in the same way and the resulting maps are shown in Fig.~\ref{fig2_8_wafermappings}, where (a) is for the 6-inch wafer and (b--h) are for a series of 4-inch wafers with different thicknesses. To better illustrate the statistical distribution, the histograms of ten wafers from Table~\ref{tab_info} are shown in Fig.~\ref{fig3_dev.mapping}(a--j). The binning width is 0.5\% and the solid lines are fitted with a Gaussian distribution~\cite{2026lorenz}, from which we obtain the standard deviation (Std.). In particular, the cyan histogram and corresponding blue line in Fig.~\ref{fig3_dev.mapping}(c) indicate the central 4-inch area of the 6-inch wafer wF. Compared with wafer wF and wI in Fig.~\ref{fig3_dev.mapping}(c) and Fig.~\ref{fig3_dev.mapping}(e) respectively, the deviation of the central 4-inch area of wafer wF shows similar results to that of wafer wI. The same trend can also be seen in Fig.~\ref{fig3_dev.mapping}(l). Since the wafers are mounted relative to a fixed standard position, it reveals that the homogeneity is similar in such an area of wafers grown in deposition device B, regardless of the wafer size. In addition, while the thicknesses of the thin films on wafers wC, wD, wF and wI are the same, the deviations strongly depend on the deposition devices, as shown in Fig.~\ref{fig3_dev.mapping}(l). Meanwhile, from the mapping result of wafers wH--wN, there exists a correlation between the thickness of deposited NbTiN films and their homogeneity of \Rs~across the wafers. As shown in Fig.~\ref{fig3_dev.mapping}(k), for wafers with thicknesses larger than 30\,nm, the Std. of the wafers approaches a constant value, and the narrow distributions of all these wafers in Fig.~\ref{fig3_dev.mapping}(d--g) indicates better homogeneity. For the wafers with film thicknesses smaller than 30\,nm, broader distributions across the wafers and the increasing Std. reveal larger inhomogeneity. This could be due to the structure-forming process of sputter-based thin-film growth. Nucleation of initial grains occurs at the beginning of sputtering, and later the islands merge, forming a continuous thin film~\cite{1998barna, 1974thornton, 2013zhang}. The initial layer of the sputtered NbTiN extends to a film thickness of approximately 20\,nm~\cite{2015zhang}. Beyond this thickness, the granular features are gradually less pronounced as the grains grow and overlap, which increase the film homogeneity.

\subsection{Temperature-dependent characterizations of superconducting properties}
Sample D1, D3 and D5 are selected to perform the low-temperature characterization. The transport measurements using the van der Pauw method are conducted to obtain \Tc. The superconducting transition is shown in Fig.~\ref{fig5_SCproperty}(a), while the inset shows the extracted \Tc~of these samples. To obtain \Tc, a Gaussian fit is used on the derivative of the temperature-dependent sheet resistance \Rsdc~as a function of temperature. We take the center of the Gaussian distribution as \Tc, and the full width of half maximum (FWHM) as the transition width indicated as the vertical bar in the inset of Fig.~\ref{fig5_SCproperty}(a). The transport measurements give the critical temperatures \Tc\textsubscript{, D1} = 13.81\,K, \Tc\textsubscript{, D3} = 13.80\,K, and \Tc\textsubscript{, D5} = 13.68\,K, and the transition widths are 0.07\,K, 0.05\,K, and 0.08\,K, respectively.

\begin{figure}
    \centering
    \includegraphics[clip, trim=0cm 0cm 0cm 0cm, width=0.99\columnwidth]{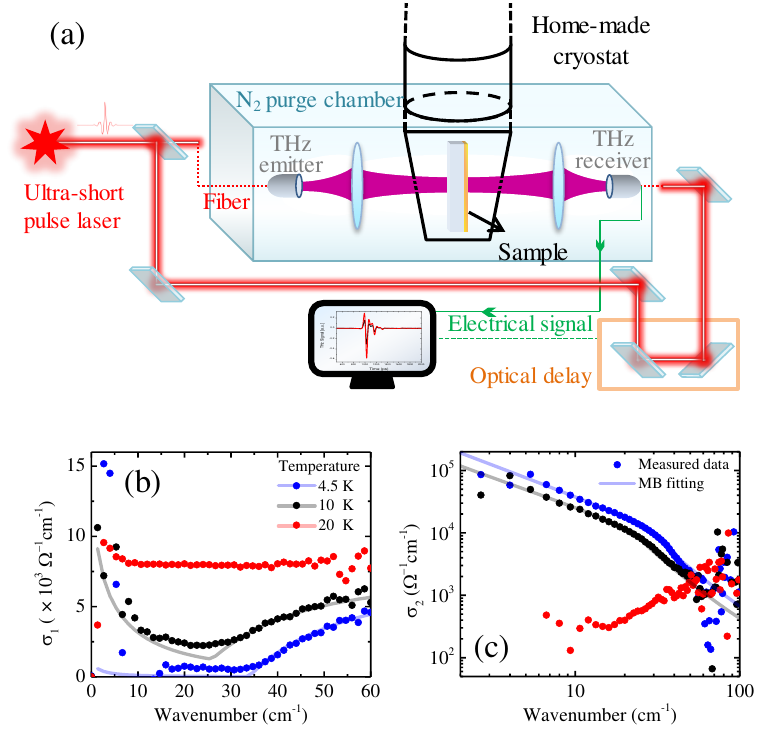}
    \caption{Overview of THz-TDS measurements. (a) THz-TDS setup with home-made cryostat. (b) Frequency-dependent complex optical conductivity of sample D3 for temperature below and above the superconducting transition (\Tc~ $\approx$ 13.8\,K). Data points are from measurements, and lines are the fits by Mattis-Bardeen (MB) theory.}
    \label{fig4_TDSsetup_sigma}
\end{figure}

\begin{figure*}[t]
    \centering
    \includegraphics[clip, trim=0cm 0cm 0cm 0cm, width=0.99\textwidth]{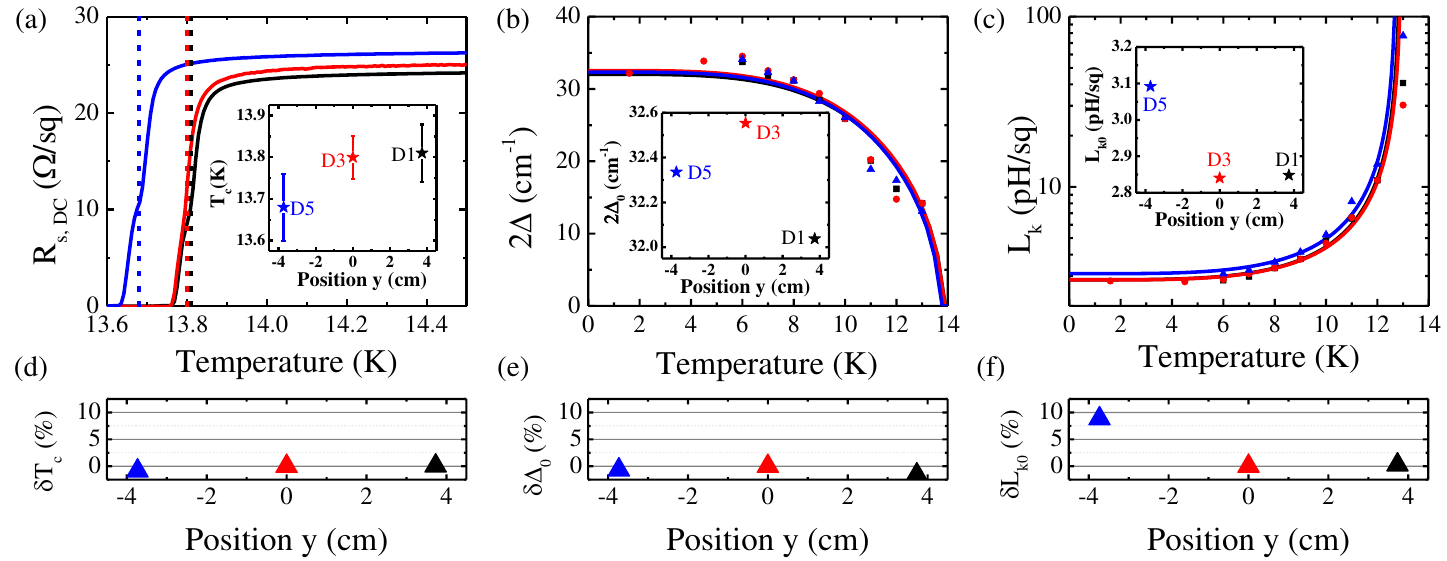}
    \caption{Comparison of superconducting properties between samples D1 (black), D3 (red) and D5 (blue). (a) DC transport measurements of \Rsdc~near the critical temperature \Tc . The inset shows \Tc~and the respective transition width as a function of position on wafer wD. (b, c) The temperature dependence of superconducting energy gap 2$\Delta$ and sheet kinetic inductance \Lk~respectively. Dots represent experimental data and the solid lines are fitting curves. The insets show the spatial distributions of 2$\Delta$\textsubscript{0} and \Lk\textsubscript{0} on wafer wD. (d--f) Position-dependent deviations of~\Tc, 2$\Delta$\textsubscript{0} and \Lk\textsubscript{0}, respectively, with respect to sample D3.}
    \label{fig5_SCproperty}
\end{figure*}

From the THz-TDS measurement, the superconducting properties can be determined via the complex optical conductivity. Fig.~\ref{fig4_TDSsetup_sigma}(b) and Fig.~\ref{fig4_TDSsetup_sigma}(c) show the real part ($\sigma_{1}$) and the imaginary part ($\sigma_{2}$) of the complex optical conductivity of sample D3, respectively, at temperatures below and above \Tc. $\sigma_{1}$ represents the absorption of the material, which decreases upon cooling due to the superconducting energy gap 2$\Delta$ opening in the superconducting state. The kink at the minimum in $\sigma_{1}$ in Fig.~\ref{fig4_TDSsetup_sigma}(b) indicates the position of 2$\Delta$, and 2$\Delta$ becomes larger as the temperature decreases. In the low-frequency range below 2$\Delta$, the thermal quasiparticle excitation gives rise to the value of $\sigma_{1}$, while above 2$\Delta$, Cooper pair breaking is observed. The $\sigma_{1}$ in the superconducting state is fitted based on Mattis-Bardeen theory~\cite{2004Tinkhambook, 1991Zimmermann} with 2$\Delta$ as the fitting parameter and with 10--55\,cm\textsuperscript{-1} as the fitting range. The fitting results are shown as solid lines in Fig.~\ref{fig4_TDSsetup_sigma}(b). The temperature-dependent superconducting gap 2$\Delta$ is obtained and shown in Fig.~\ref{fig5_SCproperty}(b). For each sample, by using the \Tc~from transport measurements, 
\begin{equation}
    2\Delta(T) = 2\Delta_{0} \sqrt{\cos\left[\frac{\pi}{2}\left(\frac{T}{T_{c}}\right)^2\right]}
    \label{eq.temp-gap}
\end{equation}
is used to extract the energy gap at zero temperature 2$\Delta_\mathrm{0}$~\cite{1966sheahen}. $T$ is the measurement temperature. The fitting curves are plotted, and 2$\Delta_\mathrm{0}$ for each sample are shown in the inset of Fig.~\ref{fig5_SCproperty}(b). 

Furthermore, in Fig.~\ref{fig4_TDSsetup_sigma}(c), the significant low-frequency increase in $\sigma_{2}$ in the superconducting state is associated with the formation of the superconducting condensate and reflects the kinetic inductive response of the superfluid. The shoulders in $\sigma_{2}$ shown in log-log plot are also related to 2$\Delta$. At low frequency, $\sigma_{2}$ is expected to exhibit 1/$\nu$ frequency dependence, with the absolute value proportional to the superfluid density \ns. However, the measured frequency range here spans the energy gap, leading to a breakdown of the 1/$\nu$ dependence for higher frequencies. To quantify the superfluid density \ns, the following relation~\cite{2016pracht} is employed:
\begin{equation}
    \label{eq.ns_lim}
    n_{s} = \frac{2\pi mc}{e^{2}}\lim_{\nu \to 0}\nu\sigma_{2}(\nu),
\end{equation}
where $m$ is the electron mass and $e$ is the electron charge. $\sigma_{2}$ is fitted by the Mattis-Bardeen theory with the same fitting parameter and range as $\sigma_{1}$, as shown in Fig.~\ref{fig4_TDSsetup_sigma}(c) solid lines. $\sigma_{2}(\nu\to0)$ is extrapolated to calculate \ns. The sheet kinetic inductance \Lk~is then derived through 
\begin{equation}
    L_{\mathrm{k}} = \frac{m}{(n_{\mathrm{s}}e^{2}d)},
    \label{eq_Lk}
\end{equation}
as shown in Fig.~\ref{fig5_SCproperty}(c). By using the two-fluid model~\cite{2004Tinkhambook}
\begin{equation}
    L_{\mathrm{k}}(T) = \frac{L_{\mathrm{k0}}}{1-\left(T/T_{\mathrm{c}}\right)^{4}},
    \label{2-fluid_Lk}
\end{equation}
the zero-temperature sheet kinetic inductance \Lk\textsubscript{0} as the fitting parameter is obtained and shown in the inset of Fig.~\ref{fig5_SCproperty}(c) for each samples.

Similar to the definition of $\delta$\Rs~in Eq.~\ref{eq.deviation}, the deviations of the superconducting properties with respect to D3 (the center piece of wafer) are calculated and shown in Fig.~\ref{fig5_SCproperty}(d--f). The deviations $\delta$\Tc~and $\delta\Delta_{0}$ are minor, while $\delta$\Lk\textsubscript{0}~is positively correlated with $\delta$\Rs~in THz-FDS across the wafer (especially D5). In the Mattis-Bardeen theory for the dirty limit $\nu\sigma_{2}(\nu\rightarrow0, T\rightarrow0) = \pi\Delta_{0}\sigma_{n}/h = \pi\Delta_{0}/(hR_{\mathrm{s}}d)$~\cite{2016pracht}, combined with Eq.~\ref{eq.ns_lim}-\ref{eq_Lk}, the \Lk~is proportional to \Rs/$\Delta_\mathrm{0}$, where $\sigma_{n}$ is the normal-state conductivity and $h$ is Planck's constant. For a 50-nm thin film demonstrated here, the variation of $\Delta_{\mathrm{0}}$ with thickness is expected to be small. Therefore, $\delta$\Lk\textsubscript{0} could be predicted by $\delta$\Rs~from the normal state. This implies that predicting the performance of superconducting devices from normal-state homogeneity could be feasible in well-understood superconducting materials, such as the present case of NbTiN.
Specifically, one could use THz room-temperature mapping of \Rs~to predict wafer-scale variations of \Lk, which could be used to adjust the design of \Lk-sensitive device according to their individual position on the wafer.

\section{Conclusion}
In summary, we demonstrate a wafer-scale mapping technique of superconducting thin films by terahertz spectroscopy. The room-temperature wafer-scale mapping reveals that for this batch of wafers, the deposition devices and thicknesses of thin films affect the thin-film homogeneity significantly, while the diameter of wafer does not. Moreover, the low-temperature measurements on sections of one wafer prove that terahertz spectroscopy is a powerful tool to characterize the superconducting properties of NbTiN thin films and their variations. Our study suggests the potential of such large-scale, contactless, and non-destructive mapping as a pre-screening tool during fabrication of superconducting quantum circuits or sensors.


\section*{Acknowledgment}
The authors would like to thank Sandra~Mebben and Gabriele~Untereiner for their assistance with sample dicing, and Ozan~Saritas, Cenk~Beydeda, and Renjith~Mathew~Roy for valuable discussion.%


\bibliographystyle{IEEEtran}
\bibliography{ref.bib}


\end{document}